\documentclass[nobalancelastpage
,prl,
 ,twocolumn
 ,secnumarabic%
,amssymb, amsmath,superscriptaddress]{revtex4}
\usepackage{epsfig}
\usepackage{amsmath}
\usepackage[T1]{fontenc}
\usepackage[latin1]{inputenc}
\usepackage{graphicx}
\usepackage{latexsym}
\usepackage{amsfonts}

\newcommand{\magn}{Fe$_3$O$_4\,$}

\newcommand{\saph}{$\alpha$-Al$_2$O$_3\,$}

\newcommand{\mfc}{$M_{\mathrm{FC}}\,$}
\newcommand{\mzfc}{$M_{\mathrm{ZFC}}\,$}
\begin{document}

\title{Finite size effects in the Verwey transition of magnetite thin films}
\author{A.M.~Bataille}
\altaffiliation{Present address: Laboratoire de Physique des mat\'eriaux, Universit\'e Henri Poincar\'e, 54506 Vand\oe uvre-les-Nancy, France}
\email{alexandre.bataille@lpm.u-nancy.fr}
\affiliation{DRECAM/SPCSI, CEA Saclay, 91191 Gif-sur-Yvette,
France}
\author{E.~Vincent}
\affiliation{DRECAM/SPEC, CEA Saclay, 91191 Gif-sur-Yvette,
France}
\author{S.~Gota}
\altaffiliation{Present address: Laboratoire Léon Brillouin, UMR
012 CEA-CNRS, CEA Saclay, F-91191 Gif-sur-Yvette, France}
\affiliation{DRECAM/SPCSI, CEA Saclay, 91191 Gif-sur-Yvette,
France}
\author{M.~Gautier-Soyer}
\affiliation{DRECAM/SPCSI, CEA Saclay, 91191 Gif-sur-Yvette,
France}

\begin{abstract}
We report on the finite size effects in the Verwey transition of
stress-free magnetite (\magn) thin films. A limit thickness of 20
nm is evidenced, above which the transition temperature $T_V$ is
constant and close to 120 K (bulk value) and below which no
genuine transition is observed. Field Cooled and Zero Field Cooled
measurements evidence irreversibilities for all thicknesses. This
irreversible behavior abruptly disappears around $T_V$ for the
thicker films, when the magnetic anisotropy vanishes. These
behaviors are interpreted in terms of assemblies of interacting
magnetic \magn clusters, which are smaller than the antiphase
domains present in the films.

\end{abstract}
\maketitle

The Verwey transition has been the subject of a huge research
effort for almost a
century\cite{review_historique_verwey_monocristaux,review_Verwey_polemique}.
The phenomenon is observed in various compounds like \magn
\cite{Nature_verwey}, Eu$_3$S$_4$ \cite{IL_polarons_Eu3S4} and
Ti$_4$O$_7$ \cite{modele_bipolarons}. Though it was first
evidenced through anomalies concerning magnetite (\magn) specific
heat\cite{premiere_anomalie_Cp} and magnetization
\cite{anomalie_magn_weiss}, the feature mostly associated to the
transition is the sharp drop in resistivity observed when heating
the sample above the transition temperature $T_V$ ($\simeq$ 120 K
for \magn). The phenomenon is extremely sensitive to the pressure
applied to the sample
\cite{R_de_T_sous_pression,R_de_T_pression_hydro} and to very
small variations of stoichiometry \cite{verwey_non_stoechio}. In a
pioneering work, Verwey carved the concept of charge ordering
\cite{Nature_verwey} in order to explain this peculiar
metal-insulator transition observed in materials in which
electronic correlations, electron-phonon coupling and kinetic
energy of the carriers are of the same order of magnitude
\cite{review_historique_verwey_monocristaux}.

The Verwey transition has given a great impulse to studies of its
prototypical material \magn, which were mostly carried out on bulk
single crystals (for recent reviews, see refs.
\cite{review_historique_verwey_monocristaux} and
\cite{review_Verwey_polemique}). However, growth and study of
\magn thin films has recently gained momentum given its potential
interest for spintronics
\cite{mon-premier-papier,pol_spin_Fe3O4_Al2O3} triggered by the
prediction of the total spin polarization (half-metallicity) of
the compound\cite{yanase}. Exploring this potential use requires
that \magn thin films be part of multilayers later patterned into
devices like spin valves and magnetic tunnel junctions
\cite{Ma_these}. Studies on the Verwey transition through
transport and magnetism measurements then provide a simple, non
destructive way of studying \magn independently of the other parts
of the device and of checking that it has not been affected during
the whole process.

From a more fundamental point of view, the study of the Verwey
transition of nanometric objects can also unravel finite size
effects. However, the studies published so far on the Verwey
transition of \magn thin films were carried out on stressed
samples given the small lattice mismatch with the substrate (MgO
\cite{margulies_PRB,MR_APB,Li_Verwey,Verwey_APB,resistivite_APB,Anomalie_MR_film,films_epaisseur_verwey,Verwey_epaisseur_Ziese},
MgAl$_2$O$_4$
\cite{Verwey_APB,Verwey_epaisseur_Ziese,Kleint_J-Phys-4} or
ZnFe$_2$O$_4$\cite{Kleint_J-Phys-4}). Epitaxy thus create an
interplay between strain and film thickness $h$, the films
adopting the lattice parameter of the substrate up to a
substrate-dependant critical thickness and relaxing continuously
afterward. The continuous decrease of $T_V$ with decreasing
thickness below $h \simeq 200$ nm has thus been ascribed to
epitaxial stress \cite{Li_Verwey}. Relaxed films are then required
in order to study the influence of the film thickness on the
Verwey transition. Besides, a superparamgnetic behavior has been
reported \cite{superpara_voogt,superpara_eerenstein} for very thin
films ($h \leq $ 5 nm), which has been ascribed to the presence of
antiphase boundaries (APBs) in the samples. These defects are
indeed observed regardless of the substrate used
\cite{margulies_PRB}, the mean antiphase domain (APD) size,
typically a few tens of nm, evolving
\cite{diffusion_APB,taille_APB_Fractales_Fe3O4} as $h^{1/2}$. A
second finite lengthscale comes into play when dealing with \magn
epitaxial thin films. The aim of this paper is to unravel finite
size effects in the Verwey transition of \magn thin films by
studying well characterized, stress-free samples, focusing on the
low thickness ($h < 50$ nm) regime.

\magn (111) thin films were grown onto \saph (0001) substrates in
a Molecular Beam Epitaxy setup dedicated to oxide thin films
elaboration by co-deposition of atomic oxygen and metal, which is
described in details elsewhere
\cite{surf-sci-Gota-Moussy,PRB_APB}. Deposition rates are
evaluated {\it in situ} using a quartz balance and also {\it ex
situ} by X-ray reflectivity ; the precision on the thickness of
the \magn layers is about 5\%. X-ray photoelectron spectra at the
Fe 2$p$ photoemission line recorded \textit{in situ} for each film
were typical of stoichiometric \magn. Punctual X-ray magnetic
circular dichroism measurement \cite{pol_spin_Fe3O4_Al2O3} also
showed no sign of deviation from perfect magnetite for the studied
film.

Full relaxation of the films is expected even for very thin films
given the large lattice mismatch (and hence the small critical
thickness) between \magn and $\alpha$-Al$_2$O$_3$ (8 \%) . This
relaxation has been evidenced by real-time RHEED (Reflection High
Energy Electron Diffraction)\cite{surf-sci-Gota-Moussy}, even
though a dilation is observed during the first growth stages . The
latter phenomenon is purely dynamic, since transmission electron
micrographs show no sign of a gradient of lattice parameter after
the end of sample growth, and all the films presented in this
paper are fully relaxed indeed. The growth is 2D as evidenced by
the RHEED patterns recorded during deposition, even though a few
films showed additional spots aside from the streaks
characteristic of 2D growth (unless otherwise stated, the RHEED
patterns of the samples considered in the following exhibited only
well defined streaks). The mean antiphase domain size does evolve
as $h^{1/2}$ and is about 25 nm for a 15-nm thick
film\cite{taille_APB_Fractales_Fe3O4}.

\begin{figure}[h]
\centering \epsfig{file=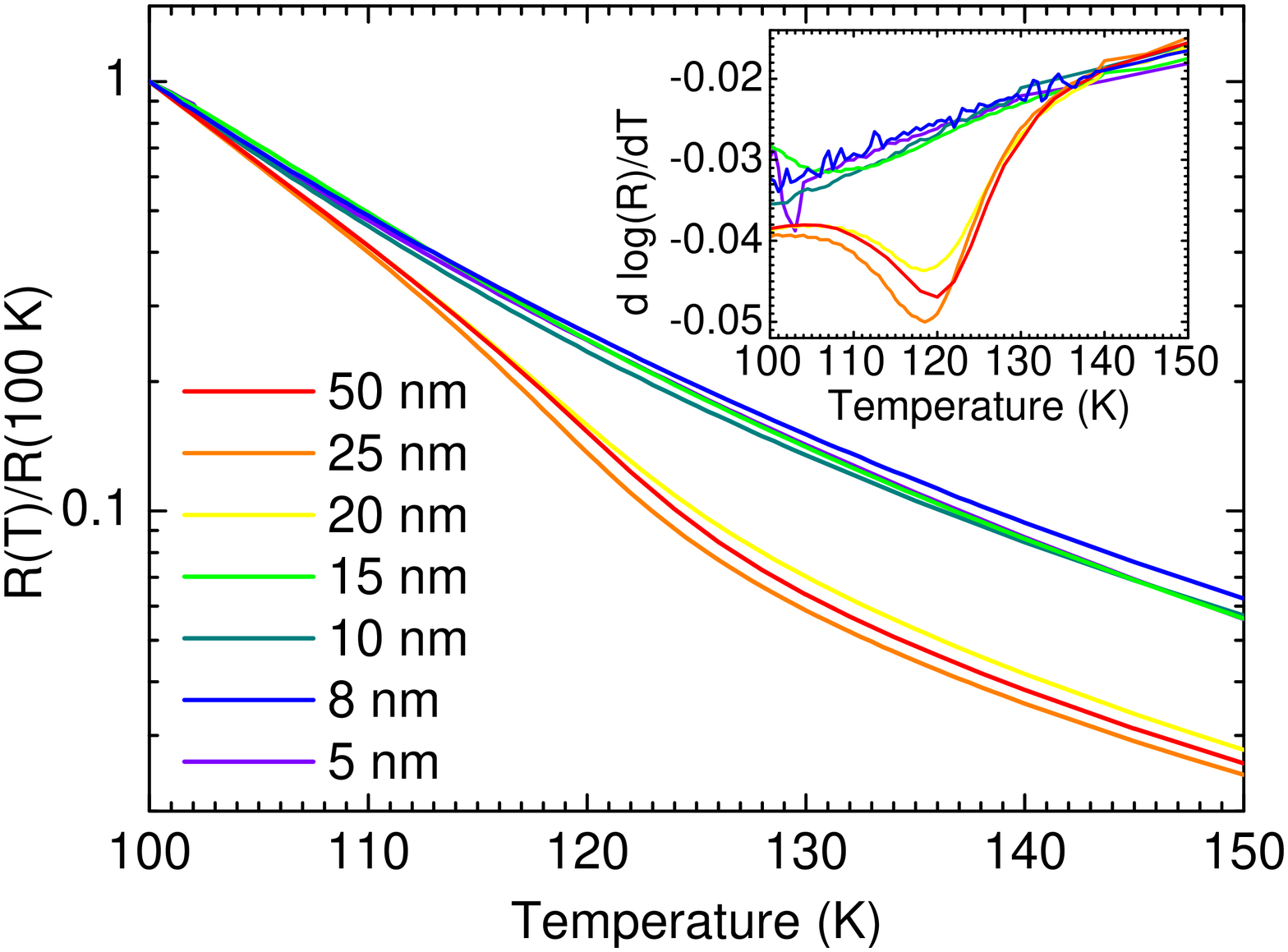,width=6.5 cm} \caption{(color
online) Transport measurements on \magn thin films. Resistivities
have been normalized with respect to $\rho(100 $K$)$ for the sake
of clarity. The inset shows the logarithmic derivative of the
resistivity, which shows the second order Verwey transition as a
minimum for films thicker than 20 nm} \label{fig:R_de_T}
\end{figure}

Two- and four-probe transport measurements were carried out within
a Quantum Design physical properties measurements system, using
either the device electronics or some external apparatus designed
for high resistance measurements (up to 10 G$\Omega$).
Measurements were performed during sample warming at a very slow
rate (less than 0.1 K/min in the transition region) in order to
ensure a complete thermalization of the sample. $R(T)$
measurements reported here were realized under zero magnetic field
though we checked that the results are not affected by the
application of constant fields up to 70 kOe.

Resistivity vs temperature measurements for seven samples are
displayed in figure \ref{fig:R_de_T}. As already reported for
stressed \magn thin films
\cite{films_epaisseur_verwey,Verwey_epaisseur_Ziese}, the Verwey
transition is second order for the \magn films grown on \saph, and
not first order as in the case of single crystals. The Verwey
temperature $T_V$ is thus defined as the minimum of the
logarithmic derivative of the resistance. In sharp contrast to the
continuous variation of $T_V$ reported for strained samples, the
films for which RHEED patterns indicated purely 2D growth fall
into two categories as a function of film thickness: those with
$h\geq$ 20 nm exhibit a clear, rather sharp Verwey transition at a
temperature close to 120 K, whereas those with $h<$ 20 nm do not
show any sign of transition. The few thick films of lesser
crystalline quality (for which RHEED patterns show spots aside
from streaks) exhibit either a reduced $T_V$ or no transition at
all.

\begin{figure}[h]
\centering \epsfig{file=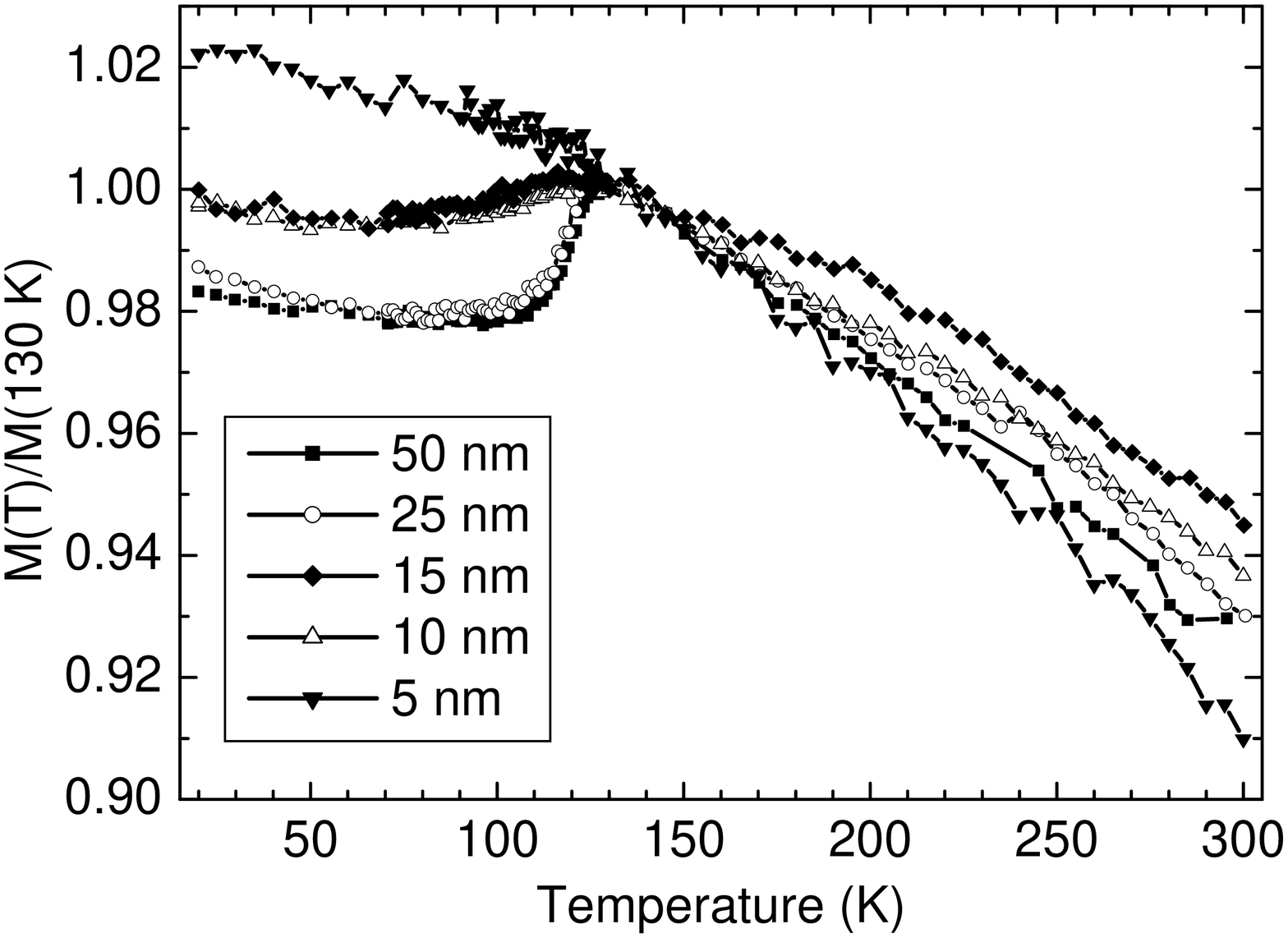,width=6.5 cm}
\caption{Field-cooled magnetization of selected \magn thin films.
Magnetizations have been normalized with respect to \mfc $(130
$K$)$ for the sake of clarity. The Verwey transition appears as a
drop of magnetization, readily seen for the 25 nm and 50 nm-thick
films. Although no genuine transition exist for the 10-nm thick
film, the \mfc curve shows a maximum at $T\simeq$ 125 K.}
\label{fig:M_de_T} 
\end{figure}

Field-Cooled (FC), together with and Zero-Field-Cooled (ZFC)
measurements were realized with a superconducting quantum
interference device (SQUID) magnetometer. The ZFC curve was
acquired first, after AC-demagnetization of the sample at room
temperature. \mfc corresponds to the reversible part of the
magnetization and FC measurements are thus the genuine test of the
Verwey transition (\mzfc and its comparison with \mfc will be
discussed later in this paper). Figure \ref{fig:M_de_T} displays
some representative examples of FC measurements. The transition
appears as a sharp variation of the FC magnetization of the 50-
and 25 nm-thick films at 119 K, $T_V$ being defined as the
temperature for which the slope of \mfc$(T)$ is maximum. The
thinnest films ($h < 10$ nm) show no signs of transition. However,
FC measurements also evidence an intermediate class of films (10
nm$\leq h\leq 15$ nm) which do not exhibit a proper transition,
but show a broad maximum of \mfc at $T\simeq$ 120 K. As for
transport measurements, $T_V$ is independent of film thickness and
close to the bulk value above a treshold thickness. Moreover, both
measurements give consistent values of this treshold thickness
above which a genuine Verwey transition is observed.


\begin{figure}[h]
\centering \epsfig{file=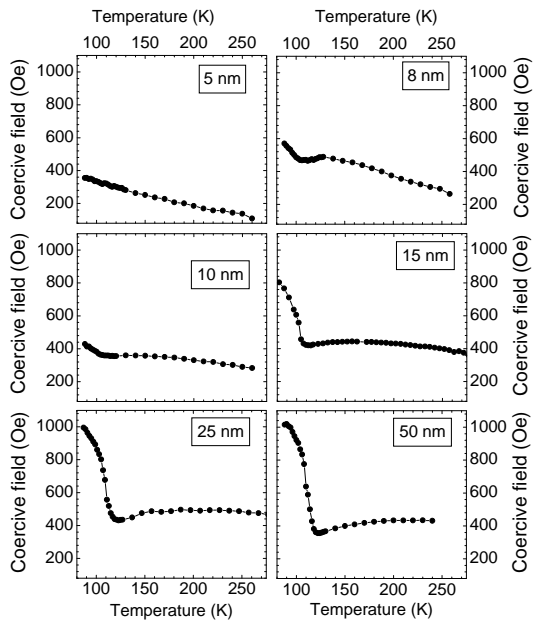,width=7 cm} \caption{Coercivity
vs temperature curves for various film
thicknesses} \label{fig:Hc} 
\end{figure}

Coercivity was measured as a function of temperature with a
vibrating sample magnetometer. A 10 kOe field was applied during
cooling to avoid twinning while cooling through $T_V$
\cite{cristallo_magnetite_BT}. No shift in the hysteresis loops
was observed, contrary to what has been reported for \magn(100)
films grown on MgO\cite{bias_APB}. 
For all but the thinner films ($h\leq 5$ nm), the coercive field
$H_c$ also presents anomalies in the vicinity of 120 K, namely a
local minimum followed in the case of the thicker films by a sharp
increase when cooling the sample below $T\simeq 120 K$ , as
evidenced by figure \ref{fig:Hc}. This phenomenon is to be linked
with the peculiar behavior of \magn anisotropy in this temperature
range. Indeed, \magn exhibits an \textit{isotropy point}, {\it
i.e.} a temperature noted $T_K$ (a few K above $T_V$) for which
the magnetic anisotropy vanishes\cite{anisotropie_magn_Verwey}.
$T_K$ can also be deduced from the $H_c(T)$ curves, since it
corresponds to the  coercivity minimum
\cite{Verwey_epaisseur_Ziese}. Fig. \ref{fig:Hc} shows that all
but the thinnest film exhibit a local minimum of $H_c$, and $T_K$
is indeed found slightly higher than $T_V$ when the Verwey
transition does occur ($h \geq$ 20 nm), and still above 110 K for
samples of intermediate thickness (8 nm $\leq h \leq$ 15 nm).

\begin{figure}[h]
\centering \epsfig{file=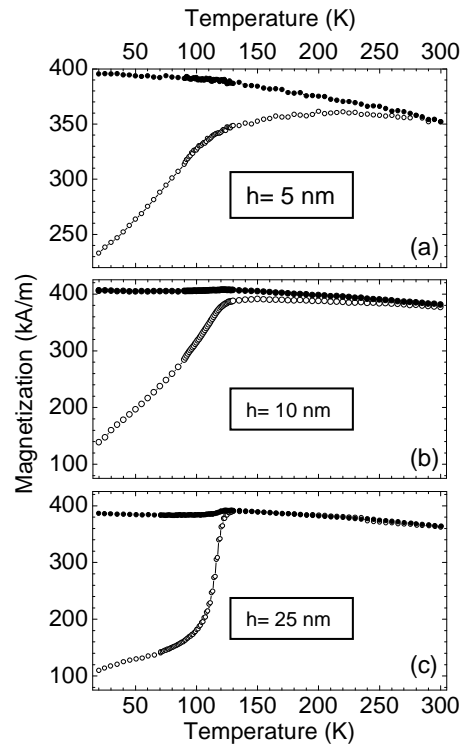,width=6 cm} \caption{Field Cooled
($-\bullet-$) and Zero Field Cooled ($-\circ-$) magnetization of
selected samples} \label{fig:FC-ZFC}
\end{figure}

We now focus on the ZFC measurements, of which relevant examples
are displayed in figure \ref{fig:FC-ZFC}. A strong separation of
the ZFC and FC curves at low temperature is observed for all
samples, and is the signature of a frozen magnetic state. In the
thinnest samples, the magnetic irreversibilities extend up to at
least room temperature (since no control of the oxygen partial
pressure was available in the magnetometer, no heating beyond 300
K was attempted to avoid the irreversible transformation of \magn
into $\alpha$-Fe$_2$O$_3$). Irreversibilities are observed in the
ZFC/FC curves of the thinnest samples for a measurement field of
1kOe which is higher than the coercive field obtained from
hysteresis loops (see Fig. \ref{fig:Hc}), suggesting that the
zero-field cooled state (established after demagnetization of the
sample) is more strongly frozen than the state obtained by
decreasing the field from the saturation value. It is likely that
the demagnetized zero-field cooled state consists in a
microstructure of numerous and strongly pinned magnetic domains.

All samples exhibit a slow increase of the ZFC curve below 100 K
(see Fig. \ref{fig:FC-ZFC}) which is characteristic of the
progressive magnetic unblocking observed in frozen
superparamagnetic systems in response to a temperature increase
\cite{dormann} , and the FC curve shows a superparamagnetic-like
decrease when heating above 120 K. However, the 25 nm film shows a
spectacular rise up of the ZFC magnetization in the 110 K region,
where an abrupt decrease of the coercive field is correlatively
observed (Fig \ref{fig:Hc}) close to the Verwey transition,
slightly below $T_K$ at which anisotropy cancels out. The frozen
magnetic domains are suddenly free to relax in this temperature
range, as would happen for frozen magnetic nanoparticles whose
anisotropy barriers decrease \cite{dormann}. This phenomenon is
only observed for thicker films ( $h$
> 20 nm), where the Verwey transition and the coercivity fall off
are well marked. The temperature dependence of the ZFC
magnetization of the thinner films (which do not exhibit these
features) is comparable with that of a frozen superparamagnet in
which some large magnetic grains remain blocked at room
temperature.

The slope of the high temperature region (150-300 K) of the FC
curves can be analyzed in terms of a superparamagnetic Curie-Weiss
behavior, in order to roughly estimate the size of the fluctuating
magnetic \magn clusters which are involved in the freezing process
at lower temperatures. Given the extremely small amount of matter
comprised in the samples, and hence the weakness of the signal, no
direct measurement of the initial susceptibility could be
performed, and we approximate it with the M/H ratio obtained from
the FC measurements under H=1 kOe.  We write:

\begin{equation}\label{eq:Curie-Weiss_clusters}
    \frac H M \simeq \frac 1\chi= \frac{3k_{\mathrm
    B}(T-\theta)}{\mu_0Nnm_{\mathrm{0}}^2}
\end{equation}

where $N$ is the density of iron atoms of moment $m_0$ in \magn,
$n$ the number of iron moments strongly coupled in a
superparamagnetic \magn cluster, and $\theta$ the Curie
temperature which corresponds to the inter-cluster interaction
energy. For the 7 samples analyzed, we obtain $\theta \simeq -700$
K, and $n$ slowly increasing from 3000 to 5000 when the film
thickness varies from 5 to 50 nm. Since the two fitting parameters
$n$ and $\theta$ depend only weakly on the film thickness, there
is no obvious relation between these magnetic entities and the
antiphase domains which are observed in \magn epitaxial thin
films. The antiphase domains are indeed 1-2 orders of magnitude
larger, and their in-plane size has been shown to vary as
$h^{1/2}$ \cite{diffusion_APB,taille_APB_Fractales_Fe3O4}, (their
volume thus varies as $h^2$). The magnetic \magn clusters which
give rise to the observed superparamagnetic behavior should be
thought of as a microstructure of the domains or of their
boundaries, the exact nature of which remains unclear.

In summary, we report on the Verwey transition on stress-free
\magn thin films epitaxially grown onto \saph as a function of the
film thickness $h$. A limiting thickness is evidenced: the films
exhibit a clear Verwey transition at a temperature $T_V$ close to
120 K (which is also the transition temperature of bulk samples)
for $h\geq 20$ nm, whereas there is no hint of transition when $h$
is below 10 nm. FC/ZFC measurements show an irreversible behavior
for all the films, the sharp variation of the ZFC curve for $h\geq
20$ nm being linked to the variation of the anisotropy, evidenced
by the anomalous behavior of the coercivity. These results are
interpreted in terms of interacting magnetic \magn clusters
smaller than antiphase domains, their size and interactions hardly
depending on film thickness.

{\bf Acknowledgments:} We would like to thank M.-J. Guittet for
her help concerning sample growth, A. Vittiglio for technical
support, D. Parker, G. Lebras, and P. Bonville for their help
during SQUID measurements, and M. Viret for fruitful discussions.


\end{document}